# Fitting Motion Models to Contextual Player Behavior

Bartholomew Spencer[1,3][0000-0001-5093-5101], Karl Jackson[2][0000-0002-7104-3054]

and Sam Robertson[1][0000-0003-8330-0011]

[1] Institute for Health & Sport (IHeS), Victoria University, Melbourne, Australia
[2] Champion Data Pty Ltd, Melbourne, Australia
[3] bartholomew.spencer@vu.edu.au

**Abstract.** The objective of this study was to incorporate contextual information into the modelling of player movements. This was achieved by combining the distributions of forthcoming passing contests that players committed to and those they did not. The resultant array measures the probability a player would commit to forthcoming contests in their vicinity. Commitment-based motion models were fit on 46220 samples of player behavior in the Australian Football League. It was found that the shape of commitment-based models differed greatly to displacement-based models for Australian footballers. Player commitment arrays were used to measure the spatial occupancy and dominance of the attacking team. The spatial characteristics of pass receivers were extracted for 2934 passes. Positional trends in passing were identified. Furthermore, passes were clustered into three components using Gaussian mixture models. Passes in the AFL are most commonly to one-on-one contests or unmarked players. Furthermore, passes were rarely greater than 25 m.

## 1 Introduction

The measurement of a player's spatial occupancy can reveal insights into space, congestion and passing opportunities. While early research into spatial occupancy considered players as fixed objects, recent iterations of Voronoi-like dominant regions have incorporated the effects of player motion [1,2]. Underlying these approaches is limited consideration of the continuous nature of space. Should the application of spatial occupancy involve possession outcomes, space should be considered relative to the ball.

Recent studies have addressed this concept. Fernandez and Bornn [3] measured the spatial dominance of teams by representing a player's influence as a bivariate normal distribution. The result considers the continuous nature of space but is not fit on empirical data. Brefeld [2] fit player motion models on the distribution of a player's observed displacements but did not consider the context of those displacements (i.e., the current possession location). Logically, the amount of spatial dominance a team exhibits over a location need be measured relative to how players would control said space if the ball were moved to that location.

In this study we present a method of fitting player motion models with consideration of displacement context. Models are fit on player commitment to passing contests, rather than raw displacements. Resultant models measure the probability a player would contest a pass to locations in their vicinity. We demonstrate the applications of these models in the analysis of kicking in the Australian Football League (AFL).

## 2 Methods

Ball tracking is not commercially available in AFL; however, ball location can be inferred from play-by-play data. Player motion models are proposed as an adequate forecast of future behaviours in the absence of precision ball tracking. Hence, the objective of this study was to model player motion with consideration of the context of player displacements, without increasing their dimensionality beyond consideration of location, velocity and time.

### 2.1 Data and Pre-Processing

LPS player-tracking data ($x, y, t$) were collected from the 2017 and 2018 AFL seasons. Tracking data (10 Hz) were consolidated with play-by-play event data (known as transactions). Transactions are recorded to the nearest second, hence are assumed to occur at the beginning of the second when combined with LPS datasets. Player orientation and velocity were calculated from the tracking data under the assumption that players were oriented in the direction of their movement. For analysis, passes that begin with and ended with a mark were extracted (*mark-to-mark* passes). This constraint ensured that location could be inferred. A *mark* is awarded when a) a player catches a kick on the full, and b) the kick travelled at minimum of 15 m.

### 2.2 Possession Contests

Commitment models are fit on player participation to forthcoming passing contests. Passing contests are pass events in which more than one player attempts to win the ball. In the AFL datasets, events that fit this criterion are *contested marks* and *spoils* transactions. The former refers to a pass caught by a player while under pressure and the latter relates to a marking attempt in which the ball is knocked away by an opponent. Passing contests are henceforth referred to as contests.

### 2.3 Modelling Process

Each contest involves two events of interest: the pass that preceded the contest and the contest transaction. The timestamps of these events are referred to as $t_p$ and $t_c$ respectively. When referring to a player's commitment we are referring to the likelihood a player will commit to a forthcoming contest, given their position and momentum at $t_p$. The commitment modelling process is as follows:

1. Player momentum and position at $t_p$ and the ball's travel time, or *time-to-point*, are recorded. The latter is simply $t_c - t_p$.
2. For each player, compute the relative location of the contest. This relative location is considered a potential player displacement. The relative location is as follows:

$$\theta = cos^{-1}\left(\frac{\overline{AB} \cdot \overline{BC}}{\|\overline{AB}\| \cdot \|\overline{BC}\|}\right) \quad (1)$$

$$(x, y) = (d \cdot \cos\theta, d \cdot \sin\theta) \quad (2)$$

where *AB* is the player's movement vector, *BC* is the displacement vector to the contest and *d* is the Euclidean distance between the player and the contest.
3. If the Euclidean distance between the player and the contest is less than two meters at $t_c$, player commitment (*C*) is recorded as 1 (hence, the player realized the potential displacement), else if greater than two meters, commitment is recorded as 0.
4. The dataset is partitioned into *commitment* and *no commitment* sets along *C*.
5. Distribution of both datasets is estimated via Kernel density estimation (KDE) with Gaussian kernels. Datasets are four-dimensional, containing the relative contest location (*x*, *y*), player velocity (*v*) and ball time-to-point (*t*).
6. The distributions are combined, weighted according to event frequency, using the following function:

$$p_i(x, y, v, t) = \frac{w f_{C=1}(x,y,v,t)}{w f_{C=1}(x,y,v,t) + (1-w) f_{C=0}(x,y,v,t)} \quad (3)$$

where $f_{C=1}$ and $f_{C=0}$ are the distributions, and *w* is the weight.

The two-meter threshold for player commitment (step 3) was chosen as an adequate distance after discussion with AFL analysts. Individual distributions represent the density of contests that were committed to ($f_{C=1}$) and those that were ignored ($f_{C=0}$). By combining the distributions (Eq. 3) the resulting variable ($p_i$) measures the probability that a new sample (given *x*, *y*, *v*, *t*) belongs to the commitment distribution. The resultant array measures a player's spatial influence. A player's influence is a forecast of their behaviors in respect to a forthcoming passing contest.

### 2.4 Spatial Metrics

We measure the spatial influence of a team as the sum of the influence of its players:

$$Inf(x, y) = \sum_{i=1}^{18} Pr_i \quad (4)$$

and dominance is the proportion of space a team owns at a location:

$$Dom_a(x, y) = \frac{Inf_a(x,y)}{Inf_a(x,y) + Inf_o(x,y)} \quad (5)$$

4### 2.5 Passing Analysis

Commitment models are used to analyse characteristics of passes. Mark-to-mark passes were extracted from the transactional dataset. The kicking distance (metres), spatial dominance, influence and *equity* of passes were recorded. AFL field equity (FE) is a measure of the value of space described in [5]. The equity of a pass is the change in FE between the passer and receiver ($equity = FE_{receiver} - FE_{passer}$). Metrics were analyzed at different field locations. Spearman correlation coefficient was used to assess the relationship between metrics and the distance between the receiver and the attacking goals. To define passing types, characteristics of passes were clustered via Gaussian mixture models, with the number of components chosen via the elbow method [6].

## 3 Results

An example output visualizing the spatial dominance and influence of an attacking team is presented in Fig. 1, where areas of darker green represent higher dominance.

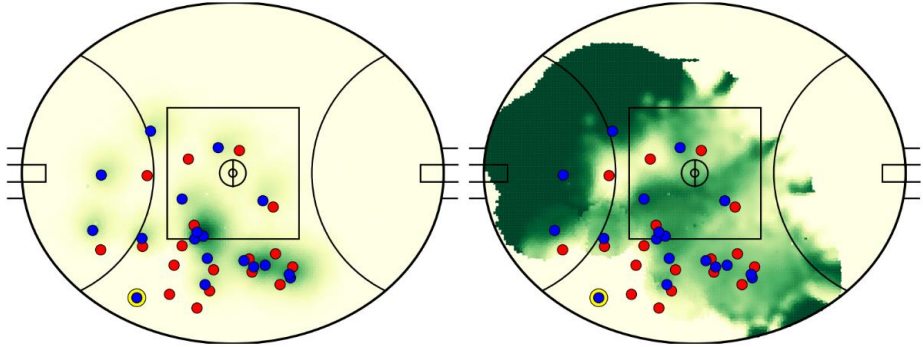

**Fig. 1.** Example output of spatial influence (left) and dominance (right) for the attacking team (in blue). The player with possession is circled in yellow (towards the lower boundary). The attacking team is moving from left to right. Spatial influence is the sum of player commitment models across a team (see Eq. 4) and dominance is a measure of spatial control derived from both team's influence (see Eq. 5). Darker green areas indicate higher values for the attacking team.

### 3.1 Commitment Models

Player commitment behavior was recorded for 46220 samples. The $C = 1$ and $C = 0$ datasets consisted of 6392 and 39828 samples ($w = 0.14$). Fig. 2 visualizes commitment models for two velocities for $t = 2$ s. These are compared to motion models fit on player displacements (as in [2]). Fitting displacements (Fig. 2b, Fig. 2d) suggests players are unlikely to reorient, hence are insufficient for modelling behavior to forthcoming contests.

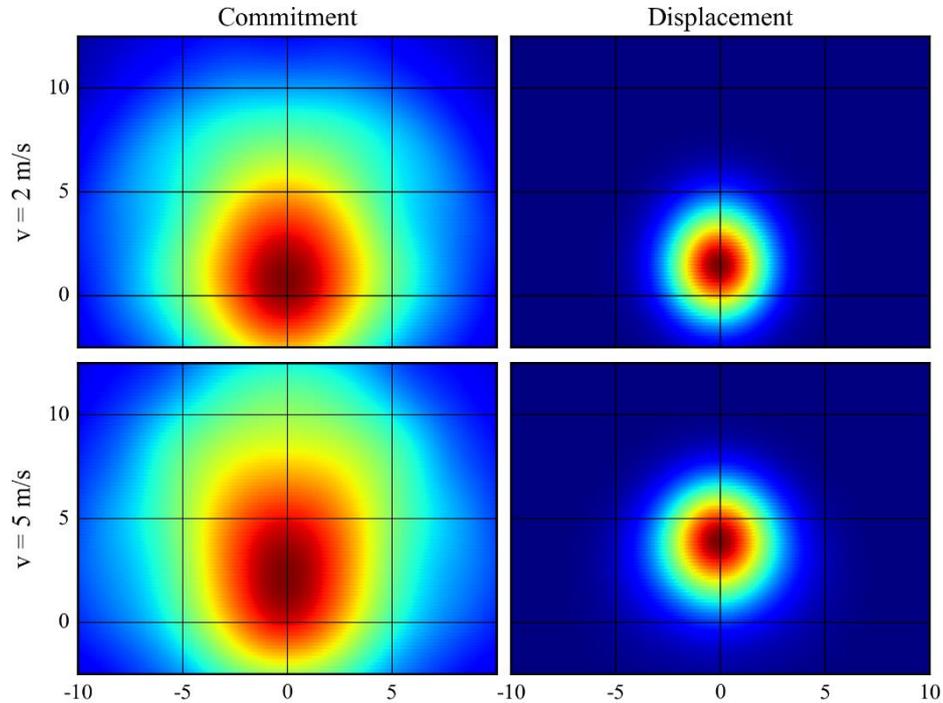

**Fig. 2.** Player commitment (left) and displacement (right) motion models for $v = 2$ m/s (top) and $v = 5$ m/s (bottom). Density represents the probability of making a displacement.

### 3.2 Passing Analysis

A total of 2934 passes were analyzed. Two-dimensional distributions of passing features are presented in Fig. 3. Dominance of passes is bimodal. The dominance and influence of receivers was recorded and smoothed by field location (Fig. 4). There is a trend towards passes to lower dominance receivers towards the attacking goal. Furthermore, influence of receivers is high in the forward 50 region. This is indicative of kicks to congested groups, rather than individual players. Minimal correlation was found between the distance to objective and both dominance ($\rho = 0.05$, $p < 0.01$) and influence ($\rho = -0.08$, $p < 0.01$).

### 3.3 Passing Clusters

Passes were clustered via GMM into three components. Component means are visualized in two-dimensions in Fig. 3. Characteristics of the components are presented in Table 1. Component 1 represents a medium-range pass to a group of players in congestion (*influence* > 0.5, *dominance* < 1.0), component 2 is a short-range pass to an open player (*dominance* = 1.0) and component 3 is a short-range pass to a one-on-one contest (*influence* < 0.5, *dominance* < 1.0).

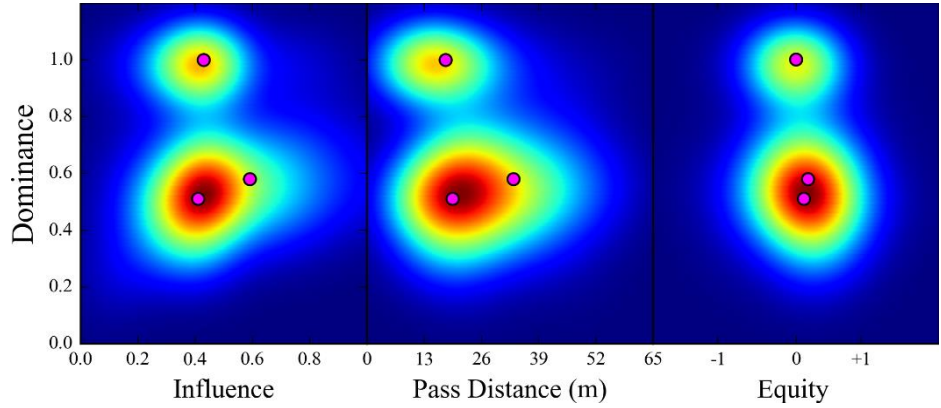

**Fig. 3.** Distributions (estimated via KDE) of (a) Influence, (b) Distance and (c) Equity relative to Dominance. GMM Component means are presented as magenta points in the 2D plots.

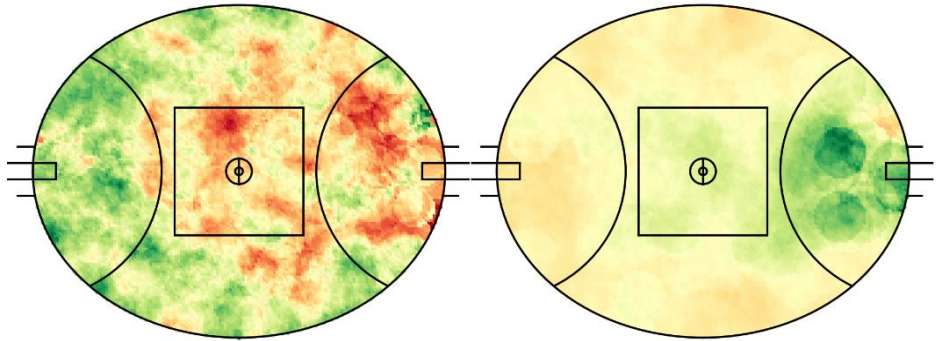

**Fig. 4.** Smoothed spatial dominance (left) and influence (right) of pass receivers. Attacking team is moving left to right. High dominance and influence is indicated by darker green regions.

**Table 1.** The weight and means of Gaussian mixture model components.

| Variable | Component 1 | Component 2 | Component 3 |
|---|---|---|---|
| Weight | 0.43 | 0.24 | 0.33 |
| Dominance (%) | 0.58 | 1.00 | 0.51 |
| Influence | 0.59 | 0.43 | 0.41 |
| Distance (m) | 33.3 | 17.9 | 19.4 |
| Equity | 0.09 | 0.00 | 0.06 |

## 4 Discussion

This study presented a method for fitting player motion models with consideration of the context of player displacements. This was achieved via the fitting of participation to forthcoming events, rather than to observed player displacements, representing a new approach to player motion models. Additionally, the models in this study fit the distribution of samples in four-dimensions, choosing to consider velocity and time as continuous rather than categorical as in [2].

It was observed that commitment models suggest a higher likelihood of reorientation than motion models fit on player displacements (see Fig. 2). In particular, displacement-based models forecast very few repositions in the negative $y$- axis. Observation of player commitment behaviors suggest reorientation is possible in all directions. The low probability of reorientation in displacement-based motion models is likely due to the nature of gameplay in AFL. The large field size and typical gameplay result in players frequently following the ball, rather than holding formations. Hence, for the analysis demonstrated in this study, motion models fit on player displacements are inadequate for describing future behavior.

Commitment models are fit on behavior to the next possession, hence are limited to applications that consider short-term behavior. At higher velocities, the spread of a player's influence increases and the shape changes (see Fig. 2). These considerations do not affect the applications presented in this study. It should be noted that commitment models were fit on 46220 samples which is roughly equivalent to the number of one-second displacements a player would make in a single match. As a result, these models may be less smooth than motion models fit on displacements (Fig. 2). Bandwidth selection during the fitting process can be modified to account for this.

A noteworthy limitation of commitment models is a reliance on transactions of differing frequency to player-tracking datasets. As a result, transactions and player-tracking may be misaligned by up to one second. The generous commitment radius of two metres deals with this to an extent, however higher frequency transactions would reduce the noise of resultant models.

Studies analyzing passing in the AFL have previously utilized discrete passing features and manually collected data (e.g., [7]). The computation of spatial features presents continuous metrics for passing analysis. Spatial dominance of receivers was found to be bimodal at dominance of an equal contest (dominance = 0.5) and an open player (dominance = 1.0). It was noted that passes to open players were rarely greater than 25 m. There is an indication that the spatial characteristics of receivers differs by region, despite minimal correlation between these metrics and a player's distance to the goalposts. In particular, the influence of receivers was higher in the forward 50 region than elsewhere. This is indicative of a pass to a congested group of players. Furthermore, early results show that receiver dominance is higher in the defensive 50 region, indicative of risk aversion in defensive positions. These results may be explained by team formations. Players have more space to work with when a team has possession in their defensive 50. This space decreases as the ball is moved towards the attacking goalposts, hence players become more congested.

Analysis of the spatial characteristics of passing produced three passing clusters. While the equity of all components was minimal, the short-range pass to an open player (component 2) had a mean equity of 0.00, hence does not typically improved a team's scoring chance. This may be a pass to stall play in the absence of better options. The

8low mean passing distance of components 2 and 3 (< 20 m) suggests a tendency to execute short-range passes.

While the analysis in this study has focused on on-ball possessions, measures of spatial occupancy have applications in off-ball analysis. Fernandez and Bornn [3] utilized similar methodology to analyze space creation of off-ball actions in soccer. Future applications of spatial occupancy should continue the development of these topics.

## 5      Conclusion

A new method for measuring player spatial occupancy was exemplified in this study. The occupancy of Australian footballers was estimated via the probability they would reposition to forthcoming passes contests. When compared to displacement-based motion models in Australian football, commitment models were found to be a better representation of contextual player behavior. Resultant commitment models were used to describe the kicking landscape of AFL footballers, finding that passes were frequently to one-on-one contests or open players. Furthermore, long kicks are infrequent and there is a significance number of passes around the minimum marking distance.

## References

1. Gudmundsson, J., Horton, M.: Spatio-temporal analysis of team sports. ACM Computing Surveys 50(2), 22 (2017). doi:10.1145/3054132
2. Brefeld, U., Lasek, J., Mair, S.: Probabilistic movement models and zones of control. Machine Learning 108(1), 127-147 (2018). doi:10.1007/s10994-018-5725-1
3. Fernandez, J., Bornn, L.: Wide open spaces: A statistical technique for measuring space creation in professional soccer. In: Sloan Sports Analytics Conference (2018).
4. Spencer, B., Morgan, S., Zeleznikow, J., Robertson, S.: Measuring player density in Australian Rules football using Gaussian mixture models. In: Complex Systems in Sport, International Congress Linking Theory and Practice, pp. 172-174. (2017).
5. Jackson, K.: Assessing player performance in Australian football using spatial data. Doctoral dissertation, PhD Thesis, Swinburne University of Technology (2016).
6. Soni Madhulatha, T.: An overview of clustering methods. IOSR Journal of Engineering 2(4), 719-725 (2012).
7. Robertson, S., Spencer, B., Back, N., Farrow, D.: A rule induction framework for the determination of representative learning design in skilled performance. Journal of Sports Sciences, 1-6 (2019). doi:10.1080/02640414.2018.1555905